\documentclass[twocolumn,floats,aps,showpacs]{revtex4}
\usepackage{times,amsmath,graphicx,latexsym}
\begin{document}

\title{Quenched Kosterlitz-Thouless Superfluid Transitions}

\author{Han-Ching Chu and Gary A. Williams}
\affiliation{Department of Physics and Astronomy, University of California,
Los Angeles, CA 90095}
\date{\today}

\begin{abstract}
Rapidly quenched Kosterlitz-Thouless (KT) superfluid 
transitions are studied by solving the Fokker-Planck equation for the vortex-pair 
dynamics in conjunction with the KT recursion relations. Power-law 
decays of the vortex density at long times are found, and the 
results are in agreement with a scaling proposal made by Minnhagen and co-workers 
for the dynamical critical exponent.  The superfluid density is strongly 
depressed after a quench, with the subsequent recovery being 
logarithmically slow for starting temperatures near T$_{KT}$.
No evidence is found of vortices being ''created'' in a rapid quench, 
there is only decay of the existing thermal vortex pairs. 
\end{abstract}

\pacs{64.60.Ht, 67.40.Fd, 67.40.Vs, 05.70.Ln}
\maketitle

There is considerable current interest in the properties of quenched 
superfluid  transitions, since they may have relevance to cosmic-string phase 
transitions in the rapidly-cooling early universe \cite{kibble}.  
Experiments \cite{mcclintock} have been 
carried out in superfluid $^{4}$He attempting to observe vortex structures 
''created'' in rapid pressure quenches through the transition, but no 
vortices have been observed, which is a puzzle \cite{he3}.

To try to understand these quenched transitions we have 
constructed \cite{ktq}
an analytic formulation of the quenched Kosterlitz-Thouless (KT) 
superfluid transition in two dimensions (2D).  Although not the 
three-dimensional (3D) realm of the experiments, the simplicity of the KT 
vortex renormalization scheme allows insights into the vortex 
dynamics that should be useful in extending the calculations to 
3D \cite{dyn}.  Previous studies 
of quenched 2D transitions have included computer 
simulations \cite{sim,luo,minn} and dynamic scaling theories 
\cite{bray,bray1,minn1}, 
but there is 
still considerable controversy over the physics involved.  The 
various dynamic scaling theories give quite 
different predictions for the temperature dependence of the dynamic 
scaling exponent $z$, and Bray and co-workers \cite{bray1} have even 
suggested that there may be 
violations of dynamic scaling.  Unfortunately the computer simulations 
are not able to resolve the questions in the field, since it has 
recently been shown \cite{minn} that different boundary conditions in 
the simulations give different results for the dynamics.

We consider a thin 
superfluid helium film on a flat substrate, which is coupled to a 
thermal reservoir at temperature $T$. The vortex pairs in the film are 
characterized by the distribution function $\Gamma (r,t)$, which is the 
density of pairs of separation between $r$ and $r+$d$r$.  It is determined from 
the 2D Fokker--Planck equation \cite{ahns},
\begin{equation}
{{\partial \,\Gamma } \over {\partial \,t}}={{2D} \over {a_o^2}}\,\;
{\partial  \over {\partial \vec r'}}\cdot \left( {{{\partial \Gamma } 
\over {\,\partial \vec r'}}+\Gamma {\partial  \over 
{\,\partial \vec r'}}\left( {{U \over {k_BT}}} \right)} \right)
\quad ,\label{1}
\end{equation}
where $r'=r/a_{o}$ with $a_{o}$ the vortex core radius, $U$ 
is the pair interaction energy, and $D$ is the diffusion coefficient 
characterizing the mutual 
friction drag force on the vortex cores of a pair.  It can be seen 
that the time in this equation is scaled by $\tau_{o}=a_{o}^{2}/2D$, 
the diffusion time of the smallest pairs of separation $a_{o}$.  In 
common with the scaling theories and simulations, we assume that any 
renormalization of D can be neglected.  At 
equilibrium where
${{\partial \,\Gamma } \over {\partial \,t}}=0$ the solution of 
Eq.\ (\ref{1}) is just the usual Boltzmann distribution,
$\Gamma =a_o^{-4}\exp \left( {-(U+2E_{c})/k_BT} \right)$ where $2E_c$ 
is the core energy of a pair.

The vortex interaction energy $U$ in Eq.\ (\ref{1}) is determined using the 
Kosterlitz-Thouless (KT) vortex renormalization methods \cite{kt}.  The 
Kosterlitz recursion relations can be written in the form
\begin{equation}
{1 \over {k_BT}}{{\partial U} \over {\partial \,r}}={{2\pi \,K} \over 
r} \label{2}
\end{equation}
and
\begin{equation}
{{\partial K} \over {\partial \kern 1pt r}}=-\kern 1pt \kern 1pt 
\kern 1pt 4\kern 1pt \,\pi ^3\,r^3K^2\,\Gamma \,\,\label{3}\; ,    
\end{equation}
where $K={\hbar ^2\sigma _s}$/${m^2k_BT}$ is the dimensionless areal 
superfluid density.  We take the core energy to be 
the Villain-model value 
$2E_c / k_BT=\pi ^2K_{o}$, where $K_{o}$ is the starting value of K at 
$r = a_{o}$.  In thermal equilibrium Eqs.\ (\ref{1}--\ref{3}) lead to the 
well-known result that the superfluid density 
has a universal jump to zero just above the transition temperature 
$T_{KT}$ \cite{kosterlitz}.

\begin{figure}[h]
\begin{center}
\includegraphics[width=0.45\textwidth]{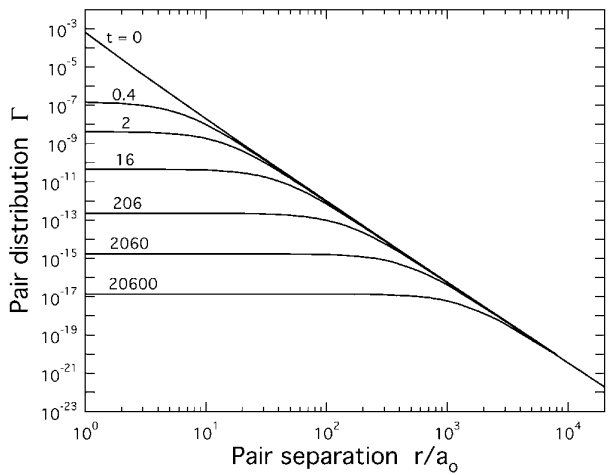}
\end{center}
\caption[]{Time dependence (in units of the diffusion time 
$\tau_{o}$)
of the pair distribution function $\Gamma$
(in units $a_{o}^{-4}$) for an 
instantaneous quench from $T_{KT}$ to 0.1 $T_{KT}$.}
\label{fig1}
\end{figure}

To study the quenched transition \cite{ktq} 
the film is first equilibrated with the heat bath at temperature $T$, 
generating an equilibrium distribution of vortex pairs.  
The top curve in Fig.\ \ref{fig1} at $t$ = 0 is the distribution at $T_{KT}$; 
it varies asymptotically as $(r/a_{o})^{-2 \pi K}$.  The temperature of 
the heat bath is then reduced 
abruptly to a low temperature, 0.1 $T_{KT}$, where in equilibrium the 
vortex density is over 20 orders of magnitude smaller.  However, due 
to the Kelvin circulation theorem the 
pairs cannot suddenly disappear, since the only way they can be 
extracted by the heat bath is if the plus-minus pairs annihilate at 
$r = a_{o}$, where the remaining core energy is converted to phonons.  This 
can take quite a long time to occur, however, since the pairs only slowly 
lose kinetic energy to the diffusive frictional force of the heat bath as 
they move together towards annihilation \cite{comment}.

To solve for the time dependence of the distribution function after 
the quench we combine Eqs.\ (\ref{1}) and (\ref{2}), which become
\begin{equation}
{{\partial \,\Gamma } \over {\partial \,t}}={{2D} \over r}\,\;{\partial  
\over {\partial r}}\left( {r{{\partial \Gamma } \over {\,\partial r}}+2\pi 
K\,\Gamma } \right) \label{4} \quad .
\end{equation}
We solve this in conjunction with Eq.\ (\ref{3}) using the method of lines
with a third-order Runge-Kutta technique on a finite domain, to a 
maximum pair separation $R/a_{o}$ = e$^{10}$ = 2.2$\times$10$^{4}$.  
The boundary conditions used are that the flux of pairs 
$J=-2D(r{{\partial \Gamma } \over {\,\partial r}}+2\pi K\,\Gamma )$ be 
continuous across the boundary $r = R$, and that $\Gamma$ drop 
abruptly to zero at $r = a_{o}$, which is equivalent to putting a 
delta-function sink term at $a_{o}$ on the right-hand side of Eq.\ (\ref{4}) to 
generate the annihilation at that point.

\begin{figure}[h]
\begin{center}
\includegraphics[width=0.45\textwidth]{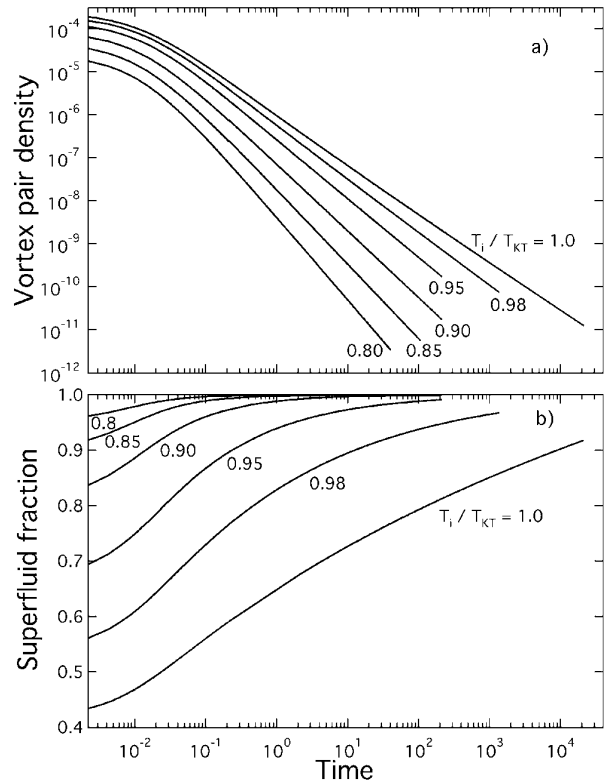}
\end{center}
\caption{a) Decay of the pair density (in units $a_{o}^{-2}$) and 
b) recovery of the superfluid fraction (evaluated at $r= R$) in time after a quench 
to 0.1 $T_{KT}$ from 
different starting temperatures $T_{i}$. }
\label{fig2}
\end{figure}

Figure \ref{fig1} shows the time dependence of the distribution function for an 
instantaneous quench to 0.1 $T_{KT}$ from a starting temperature
barely below $T_{KT}$ (from $K_{o}$ = 0.747853, compared to the 
critical value 0.747852 at $T_{KT}$).  As expected for the 
annihilation process the smallest 
pairs decay away first, in the classic pattern of a phase-ordering 
transition where at long times the largest scales become more 
dominant.  By integrating  the distribution function over d$^{2}r$ the 
pair density is obtained, shown as the top curve in Fig.\ \ref{fig2}a. 
At long times the density decreases nearly as $t^{-1}$, in agreement 
with the scaling theories \cite{bray,bray1,minn1} at $T_{KT}$.
We find, however, that the slope only approaches -1 logarithmically 
slowly, being -1.12 at $t$ = 100 and -1.09 at $t$ = 
2$\times$10$^{4}$, and more detailed fits show that the density is 
varying as 
$(t\ln t)^{-1}$.  
This is a consequence of the slow recovery of the 
superfluid density after the quench (bottom curve of 
Fig.~\ref{fig2}b ), which has a logarithmic approach to its 
equilibrium value. The substantial drop in the superfluid 
fraction just after the quench (from an initial value 0.89)
is due to the increased polarizability
of the pairs at low temperature.  Their density is unchanged 
immediately after the quench, but they screen more effectively due to 
the lack of thermal fluctuations. 

\begin{figure}[h]
\begin{center}
\includegraphics[width=0.4\textwidth]{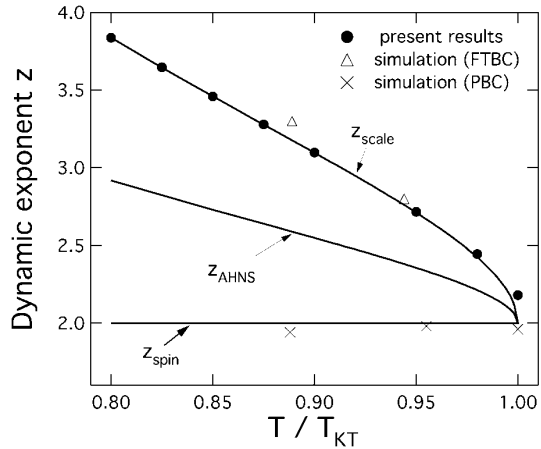}
\end{center}
\caption{Dynamic exponent z calculated from the slope of the curves 
in Fig.\ \ref{fig2}a, compared with various scaling theories.  The PBC simulations 
are from Ref.\ \cite{luo}, the FTBC from the relaxational dynamics of 
Ref.\ \cite{minn}.}
\label{fig3}
\end{figure}

Quenches from starting temperatures below $T_{KT}$ also display 
power-law variation of the vortex density, since the KT transition is 
a line of critical points.  The plots in Fig.\ \ref{fig2} show this 
for a series of initial temperatures between 0.8 $T_{KT}$ and $T_{KT}$.
At the lower starting temperatures the superfluid density recovers more 
quickly, and the vortex decay then varies accurately with the form $t^{-z/2}$,
where the constant $z$ agrees with the scaling prediction of Minnhagen and co-workers 
for the dynamic exponent \cite{minn1},
\begin{equation}
z_{scale}=4{{\sigma _s(T)} \over {\sigma _s(T_{KT})}}{{T_{KT}} \over T}-2
\quad ,  \label{5}
\end{equation}
Figure \ref{fig3} shows the values of $z$ extracted from the slopes of 
the curves in Fig.\ \ref{fig2}a, which are compared with $z_{scale}$, with a 
dynamic exponent predicted in Ref.~\cite{ahns}, and with  
simulation results.  It  
has recently been shown \cite{minn} that the simulations 
of the vortex dynamics can be 
misleading, since different results for $z$ are obtained depending on 
the boundary conditions used:  periodic boundary conditions (PBC) 
yield \cite{luo}
only the spin-wave result $z$ = 2 at all temperatures, 
while fluctuating-twist boundary conditions 
(FTBC) give $z_{scale}$ \cite{minn}. Our results in agreement with $z_{scale}$ 
(except near $T_{KT}$ where the logarithmic corrections enter) 
provide further support that this is the correct exponent characterizing the 
vortex dynamics.  Unlike the simulations our formulation does not 
contain spin waves, and 
this allows a direct probe of the vortices, without their behavior
being obscured by the slower dynamics of the spin waves.

Bray and co-workers \cite{bray1} have predicted
the appearance of a logarithmic term in the decaying vortex density for 
quenches starting above $T_{KT}$,
but with a form $\ln t/t$ that differs from ours.  
In their calculation they assume, however, that
the superfluid density is a constant in time.  
As seen in Fig.~\ref{fig2}b this may be a reasonable 
approximation for quenches starting well below $T_{KT}$, but it 
appears to fail near $T_{KT}$ because 
of the logarithmic recovery of the superfluid density. The 
log term in our vortex density arises entirely from this variation. In 
Ref.\ \cite{bray1}
the appearance of the logarithm is ascribed solely to the presence 
of free vortices, and  this is cited as a violation of dynamic scaling.  
This is not the case in our calculation, where no free vortices at 
all are present.  It appears that the development of the log term 
only depends on a sufficient density of pairs of large separation; for quenches 
starting from $T/ T_{KT}$ = 0.98 we find that the deviations from the 
power-law decay are already strong, but not quite the logarithmic behavior 
found right at $T_{KT}$. 

\begin{figure}[h]
\begin{center}
\includegraphics[width=0.45\textwidth]{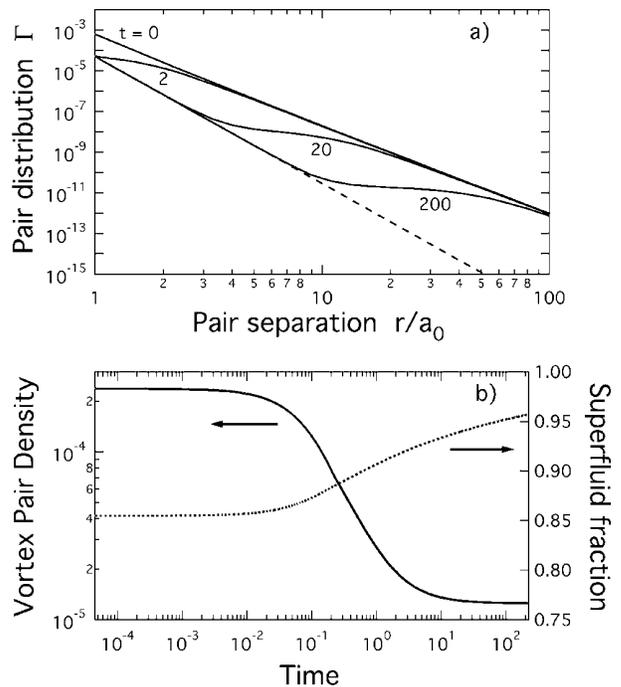}
\end{center}
\caption{a) Pair distribution function and b) vortex density and 
superfluid fraction (dotted curve) versus time for a quench
from $T_{KT}$ to 0.75 $T_{KT}$.}
\label{fig4}
\end{figure}

Figure \ref{fig4} illustrates the approach to equilibrium following an 
instantaneous quench from $T_{KT}$ to 0.75 $T_{KT}$.  In the 
distribution function the smallest 
pairs quickly come to thermal equilibrium (the dashed
line) at the new temperature, while the larger pairs only slowly 
readjust.  The vortex density levels 
off near the equilibrium value, since it is dominated by the smaller 
pairs. The superfluid density, determined by the larger pairs, recovers 
more slowly, although because of the higher temperature its initial 
reduction is considerably smaller than that in Fig.\ \ref{fig2}b.

\begin{figure}[h]
\begin{center}
\includegraphics[width=0.45\textwidth]{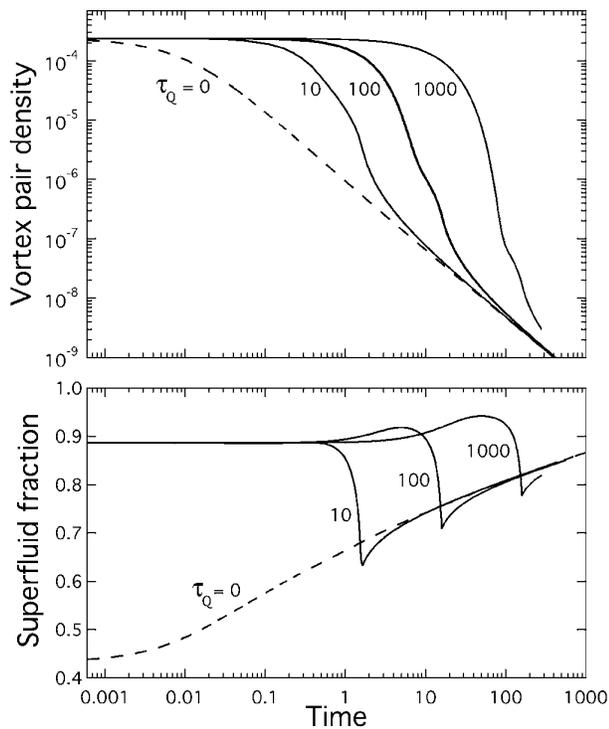}
\end{center}
\caption{Vortex density and superfluid fraction versus time for quenches 
from $T_{KT}$ to 0.1 $T_{KT}$
in time $\tau_{Q}$ (in units of the diffusion time).}
\label{fig5}
\end{figure}

From Fig.\ \ref{fig2}a it is clear that no vortices are being ''created'' in  
the instantaneous quenches from $T_{KT}$ and below; there is only 
decay of the existing thermal vorticity.  We have looked at the effect 
of the quench time on this decay, shown in Fig.\ \ref{fig5}.  This shows the 
vortex density and superfluid fraction for the case of a linear ramp of 
the temperature from 
$T_{KT}$ to 0.1 $T_{KT}$ in a time $\tau_{Q}$.  We assume for 
convenience that D is independent of temperature, although in a real 
helium film it would vary during such a finite-time quench \cite{heD}.
It can be seen in Fig.\ \ref{fig5}a that for a finite quench time 
the vortex density decreases more slowly than for the instantaneous 
quench (dashed curve), since the system spends more time at higher 
temperatures where the thermal density is higher.  
As the temperature falls and the smaller pairs that had been able to 
stay in quasi-equilibrium start to annihilate, the more sluggish 
larger pairs begin to dominate, and the curves
revert to the instantaneous-quench result at long times. The slight 
wiggles in the curves are due to changes in the superfluid fraction,
shown in Fig.\ \ref{fig5}b.  The fraction begins to increase slightly as 
the density of the smaller quasi-equilibrium pairs decreases with
the decreasing temperature, but finally the increasing polarizability 
of the larger pairs overcomes this, driving down the superfluid 
fraction.  Since there are now more pairs remaining than in the 
instantaneous quench the fraction drops even below the value for that 
case, and then rejoins it at long times as the smaller pairs decay 
completely and the larger pairs dominate. 

The behavior of the vortex density in Fig.\ \ref{fig5} is the opposite of the 
Kibble-Zurek predictions of Ref.\ \cite{kibble}, which postulate that the 
faster the quench is performed the more vortices 
will be created.  Since these theories do not take into account the 
thermal vortices it is unclear how any extra vortices 
would be formed.  Of course, our calculations cannot be extended to 
starting temperatures above $T_{KT}$, since the KT recursion relations 
become invalid.  We doubt, however, that the additional thermal 
vorticity generated at higher temperatures will do anything but 
add smoothly to the decaying curves of Fig.\ \ref{fig5}.  The 2D 
quench simulations from above $T_{KT}$ \cite{sim} show no 
evidence of any additional vortex 
creation.  Even simulations of 
quenched 3D transitions show (in Fig.\ 
\ref{fig1} of Ref.\ \cite{antunes})
only a smooth and monotonic decay of the thermal vorticity 
starting from above $T_{c}$,
with no extra vorticity being generated.

In summary, an analytic formulation of 2D quenched superfluid transitions 
shows a power-law decay of the vortex density with time, 
in agreement with the predictions for the dynamic scaling exponent
of Minnhagen and co-workers.  Quenches starting near $T_{KT}$ show an 
additional logarithmic variation resulting from the slow recovery of 
the superfluid density to its equilibrium value.  The strong depression of the 
superfluid density following a quench has not been previously 
been taken into account, though it plays a major role in the quench 
dynamics, as seen in Fig.\ \ref{fig5}.  It will be important that any future 
simulations of quenched transitions (in both 2D and 3D) include computations 
of the superfluid fraction (the helicity modulus) as well as the vortex density, 
and also study the 
effect of changing the boundary conditions.  Experimental tests of 
these vortex dynamics should also be possible:  measurements on 
Josephson-junction arrays have been able to determine $z$ = 1.98$\pm$ 
0.03 at $T_{KT}$ \cite{clarke}, and experiments on liquid-crystal films 
are able to visualize the pair-defect annihilation process following a 
quench \cite{clark}.

This work is supported by the National Science Foundation, DMR 97-31523.


\begin{thebibliography}{25}
\bibitem{comment}
In this formulation it is not necessary to follow the complex motion and 
reconnection of the pairs with each other, as such motion is reflected 
statistically in the change of the distribution function after each 
time step.  Our recombination process at $a_{o}$ is the same as the 
``bulk recombination'' process discussed by other authors who follow 
the collision of individual pairs \cite{ahns}.
\end{thebibliography}
\end{document}